\documentstyle[12pt]{article}
\footskip- 15mm
\begin{document}
The seminal papers by Pecora and Carrol (PC) [1]  and Ott, Grebogi 
and Yorke (OGY) [2] in 1990 have induced avalanche of research works 
in the field of chaos control. Chaos synchronization in dynamical 
systems is one of methods of controling chaos, see, e.g. [1-8] and 
references therein.The interest to chaos synchronization in part is 
due to the application of this phenomenen in secure communications, 
in modeling of brain activity and recognition processes,etc [1-8]. 
Also it should be mentioned that this method of chaos control may 
result in improved performance of chaotic systems [1-8].
According to PC [1] synchronization of two systems occurs when the 
trajectories of one of the systems will converge to the same values as 
the other and they will remain in step with each other. For the 
chaotic systems synchronization is performed by the linking of 
chaotic systems with a common signal or signals (the so-called 
drivers): suppose that we have a chaotic dynamical system of 
three or more state variables. In the above mentioned way of chaos 
control one or some of these state variables can be used as an input 
to drive a subsystem consisting of remaining state variables and which 
is a replica of part of the original system.In [1] it has been shown 
that if the real parts of the Lyapunov exponents for the subsystem 
(below: sub-Lyapunov exponents) are negative then the subsystem 
synchronizes to the chaotic evolution of original system. If the 
largest sub-Lyapunov exponent is not negative, then one can use the 
nonreplica approach to chaos synchronization [9]. Within the 
nonreplica approach to chaos synchronization one can try to 
perform chaos synchronization between the original chaotic system and 
nonreplica response system with control terms vanishable upon 
synchronization. To be more specific, one can try to make negative the 
real parts of the conditional Lyapunov exponents of the nonreplica 
response system. As it has been shown in [9] from the application 
viewpoint nonreplica approach has some advantages over the replica 
one. \\ 
Recently in [10] it has been indicated that for more secure
communication purposes the use of hyperchaos is more reliable. Quite 
naturally in the light of this result the investigation of hyperchaos
synchronization is of paramount importance. According to Pyragas for 
hyperchaos synchronization at least two drive variables are needed 
[11].
Recently this idea was challenged in [12] in the sense that instead 
of several driving variables one can try to drive the response system
with a scalar combination of those driving variables. But one should 
keep in mind that in this case the synchronization occurs between the 
nonreplica system and original chaotic system. Recent paper [13] also 
falls into this category, although its authors are using only single 
control term added to the replica response system.In recent work [14] 
the classification of different types of synchronization is
conducted. Such a classifiation into different types corresponds to 
the different values for the sub- (or conditional) Lyapunov exponents 
and still there is no unique generally accepted classification. For 
example, according to [15] if one of sub-Lyapunov exponents is equal 
to zero, while others are negative, then one can still speak of 
synchronization between the response and drive systems in the general 
sense: a generalized synchronization introduced for drive-response 
systems is defined as the presence of some functional relation 
between the states of response and drive. According to [14], the 
similar situation could be characterized by the so-called
marginal synchronization:there are there types of marginal 
synchronization: 1) marginal constant synchronization: in this case 
the response system becomes synchronized with the drive, but with a 
constant separation.\\
2) marginal oscillatory synchronization: this type of synchronization 
implies that the difference between the drive and response will change 
in an oscillatory fashion with a frequency that will depend on the 
imaginary part and with constant amplitude that will be related to 
the difference at the moment in which the connection starts.\\
3) sized synchronization: in this type of synchronization also one 
has a single zero sub-Lyapunov exponent; in this case the observed 
behavior is different from the case of marginal constant 
synchronization and consists in that the response system exhibits the 
same qualitative behavior as the drive, but with different size (and 
sometimes with different symmetry); as a prominent example of this 
type of synchronization one can cite the case $z$ driving for the 
classical Lorenz model [7]. It is easy to show that  one of sub-
Lyapunov exponents for the Lorenz model in the case of $z$ driving 
really is equal to zero. Really, writing the corresponding equation 
for the sub-Lyapunov exponents and taking into account the results of 
[16], where it has been shown that for those dynamical systems, whose 
chaotic behavior has arisen out of instability of the steady state 
solutions (fixed points) while calculating the sub-Lyapunov exponents 
one can replace the time dependent solutions of the dynamical systems 
with the steady state (st) solutions, we obtain that  for the case of 
$z$ driving one of the sub-Lyapunov exponents is equal to zeo.\\
In the above mentioned papers [14-15] the presented examples represent
third-order nonlinear dynamical systems.\\
In recent communication we have presented an example of marginal or 
general type synchronization in one of high (four) dimensional 
R\"ossler systems with a single driving variable [17].\\
As it has been indicated in [18], in principle hyperchaos 
synchronization could be performed even with a zero driving variable. 
The main idea behind this work is to make negative sub-(or 
conditional) Lyapunov exponents negative by the changing of system's 
parameters. Although this idea was put forward by the authors of [18], 
the concrete example of the application of the idea to the specific 
hyperchaos systems is not presented yet.\\
With this paper we fill in this gap.\\
So, in this communication we present the first (to our knowledge) 
example of hyperchaos synchronization with a zero driving variable in 
one of R\"ossler models. The system under consideration is of the form 
[13]: $$\frac{dx}{dt}= - y - z , $$
$$\frac{dy}{dt}= d x+ a y + w, \hspace*{3cm}(1)$$
$$\frac{dz}{dt}= 3 + xz, $$
$$\frac{dw}{dt}= c z + b w, $$
According to [13], nonlinear system (1) exhibits hyperchaotic behavior
with $a=0.25, c=-0.5, b=0.05, d=1$.
Acting along with the algorithm proposed in [18], we adjust these 
parameters in the following manner:
$$ a_{1}=a-\epsilon_{1} (y-y_{g}),b_{1}=b-\epsilon_{2} (w-w_{g}),$$
$$ c_{1}=c-\epsilon_{3} (z-z_{g}), d_{1}$$
$$=d-\epsilon_{4} (x-x_{g}),\hspace*{2cm}(2)$$
where $\epsilon_{1,2,3,4}$ are the control coefficients; $x,y,z,w$ 
describes the orbits of the response system; $x_{g}, y_{g}, z_{g}, 
w_{g}$ are the goal orbits. When synchronization is implemented, 
$x=x_{g}, y=y_{g}, z=z_{g}, w=w_{g}$.
The aim is to change the system's parameters so that to make negative 
the roots (or the real parts of these roots) of the characterisic 
equation, which follows from the procedure of finding of the 
eigenvalues of the Jacobian corresponding to the response system (1). 
(Let us remind that in the case of a zero driving, the 
response system coincides with the original nonlinear system (1) 
[18]).\\ 
$$\lambda (\lambda -s_{1})(\lambda -x_{g})(\lambda -s_{2})$$
$$+s_{4}(\lambda -x_{g})(\lambda -s_{2}) + z_{g}s_{3}$$
$$+z_{g}(\lambda -s_{1})(\lambda -s_{2})=0,\hspace*{3cm}(3)$$
where
$$s_{1}=a-\epsilon_{1} y_{g}, s_{2}=b-\epsilon_{2} w_{g},$$
$$ s_{3}=c-\epsilon_{3} z_{g}, s_{4}=d$$
$$-\epsilon_{4} x_{g},\hspace*{2cm}(4)$$
As our task is to make negative the roots of the equation (4), we can 
try different possibilities.\\
First of all for $s_{3}=s_{4}=0$ the roots of the equation (4) 
satisfy the following equality:
$$(\lambda -s_{1})(\lambda -s_{2})(\lambda^{2} -x_{g}\lambda $$  
$$+z_{g})=0,\hspace*{2cm}(5)$$
It is quite easy to see that by choosing $\epsilon_{1,2}$ and using 
the results of Pyragas [11] about the negativity of the roots of the 
quadratic equation in (5) in the case of $(x,y,z,w)_{g}$ being the 
solutions of the initial R\"ossler model (1) these roots could be done 
negative.\\
In the case of $s_{3}=0$ the equation (3) could be rewritten in the 
following form:
$$(\lambda -s_{2})(\lambda (\lambda -s_{1})(\lambda -x_{g})$$
$$+s_{4}(\lambda -x_{g})+z_{g}(\lambda -s_{1}))=0,\hspace*{3cm}(6)$$
In other words, one of the roots could be done negative by choosing 
of the value of $\epsilon_{2}$:the sign of $b-\epsilon_{2} w_{g}$ 
should be negative. As it has been shown in [18], larger values of 
parameter changes could be avoided by allowing the response system run 
freely accepting $\epsilon =0$.
The three remaining roots of (6) satisfy the cubic equation: 
the negativity condition for the roots of the cubic equation could be 
written by Routh-Hurwitz criteria:
$$f_{1}=-(s_{1}+x_{g})>0, f_{3}=-s_{4}x_{g}-z_{g}s_{1}>0,$$ 
$$f_{4}=f_{1}f_{2}-f_{3}>0, \hspace*{2cm}(7)$$
where
$$f_{2}=s_{1}x_{g} + s_{4} + z_{g},\hspace*{5cm}(8)$$
and $f_{1,2,3}$ are the coefficients before $\lambda^{2,1.0}$ 
respectively.As an example of demonstration of obtaining negative 
values for the roots of cubic equation take for simplicity $x_{g}=0$.
As $s_{3}$ is dependent on $\epsilon_{3}$, this can be 
done quite easily for the given $z_{g}$. Then from (4), (7) and (8) 
we establish that $s_{4}=1, s_{1}<0$. It is quite easy to establish 
from (7) in this case synchronization occurs if $z_{g}>0$. So in the 
concrete case we have demonstrated that in principle it is possible 
to synchronize hyperchaos with zero driving variable by changing the 
system's parameters. For the concrete example of the R\"ossler model 
in more general case let us write the Routh-Hurwitz criteria 
explicitly. First write the characteristic equation to be investigated 
in the form
$$\lambda^{4} + m_{1}\lambda^{3} + m_{2}\lambda^{2}$$
$$ + m_{3}\lambda + m_{4}=0,\hspace*{2cm}(9)$$ 
where
$$m_{1}=-(s_{1} + s_{2} + x_{g}),$$ 
$$m_{2}=s_{1}s_{2}+x_{g}(s_{1}+s_{2}) +s_{4}+z_{g}),$$
$$m_{3}=-(x_{g}s_{1}s_{2}+s_{4}(x_{g}+s_{2})+z_{g}(s_{1}+s_{2})),$$
$$m_{4}=s_{2}s_{4}x_{g} + z_{g}s_{3}$$  
$$ +z_{g}s_{1}s_{2},\hspace*{3cm}(10)$$
Now the conditions for the negative roots could be written in the 
form:
$$m_{1}>0, m_{1}m_{2}-m_{3}>0, m_{1}m_{2}m_{3}-m_{3}^{2}$$
$$-m_{4}m_{1}^{2}>0, m_{4}>0,\hspace*{1cm}(12)$$ 
So for the given $x_{g},y_{g},z_{g},w_{g}$ we have for arbitrary 
parameters $\epsilon_{1,2,3,4}$ by choosing which one can try to make 
negative the real parts of the roots of equation (10).Larger values of 
$\epsilon$ could be avoided by allowing free running for the response 
system by taking the corresponding $\epsilon$ as zero.\\
From the application point of view the obtaining of the negative
values for the sub-(or conditional) Lyapunov exponents are very 
important. As the synchronization time is inversely proportional to 
the largest Lyapunov exponent, it is quite essential to make these 
exponents larger in magnitude. As it seems to us for this purpose the 
combination of the nonreplica approach to chaos synchronization [9] 
(see, also [19-21]) with the method of parameter changes could be 
quite appropriate, as in the case of nonreplica approach the 
nonreplica response system will include the additional arbitrary
parameters by choosing which one can achive the desired goal.\\
In this connection it is worth while to study hyperchaos 
synchronization in the case of a single driving. As it is clear from 
the results of [13], hyperchaos synchronization in the R\"ossler model 
under investigation would be possible within the nonreplica approach 
if the single arbitrary control parameter exceeds the threshold value. 
By applying the method of parameter changes with combination with the 
nonreplica approach one can obtain that the threshold value could be 
lowered, or even be eliminated. These investigations will be
presented elsewhere in more detail.\\
For the first time an example of hyperchaos synchronization with 
a zero driving variable using the system's parameters changes, 
proposed by L.Zonghua and C.Shigang (Phys.Rev.E, {\bf55}, 6651 (1997)) 
is presented. Also it has been indicated that in the case of a single 
variable driving put forward by A.Tamasevicius and A.Cenys, 
Phys.Rev.E, {\bf55}, 297 (1997) such an approach allows one to avoid 
the threshold effect on the value of the parameter responsible for 
the "weight" (intensity) of nonreplica approach to the chaos 
synchronization proposed by M.Ding and E.Ott, Phys.Rev.E,{\bf49}, 
R945 (1994).\\ 

\end{document}